\documentclass[11pt,a4paper]{article}
\usepackage[utf8]{inputenc}
\usepackage{amsmath}
\usepackage{graphicx}
\usepackage{authblk}
\usepackage{url}
\usepackage{geometry}
\usepackage{siunitx}
\usepackage{subcaption}  
\usepackage{pdflscape} 
\usepackage{newunicodechar}
\usepackage{hyperref}
\newunicodechar{₂}{$_2$}
\DeclareUnicodeCharacter{202F}{\,}  

\title{\textbf{Thermal Cesium Release in PMTs Revealed by Resonant Laser Spectroscopy and Its Correlation with Quantum Efficiency}}
\author[a,b]{A. De Benedittis}
\author[a]{P. Migliozzi}
\author[a]{C.M. Mollo}
\author[a]{A. Simonelli \footnote{Corresponding author [andreino.simonelli@na.infn.it]}}
\affil[a]{INFN Sezione di Napoli, Laboratorio CAPACITY, Caserta, Italy}
\affil[b]{Dipartimento di Matematica e Fisica, Università degli Studi della Campania “Luigi Vanvitelli”, viale Lincoln 5, Caserta, 81100, Italy}
\date{}

\begin{document}

\maketitle

\begin{abstract}
We report optical evidence of cesium (Cs) evaporation from a bialkali (SbKCs) photocathode during controlled heating of a photomultiplier tube (PMT). A DFB laser scanned across the 852.113 nm Cs D2 line reveals absorption features only above \SI{60}{\celsius}, indicating thermal desorption. The absorption correlates with temperature and offers a non-invasive method to monitor photocathode degradation in sealed detectors.
\end{abstract}

\section*{Introduction}
The stability and quantum efficiency (QE) of bialkali photocathodes are critical parameters for the performance of photomultiplier tubes (PMTs), especially in low-background applications such as neutrino detection and dark matter searches. Among the alkali metals used, cesium (Cs) plays a dominant role in determining the spectral sensitivity and surface work function of the photocathode. Under thermal or operational stress, alkali metals may desorb from the internal surfaces of the PMT, leading to QE degradation and altered charge transport. Over the last decade, significant efforts have been made to improve the characterization of PMT photo-response, including the development of scanning setups for spatial QE mapping~\cite{set}, timing diagnostic instruments~\cite{Bozza:2016dtf}, and large-scale qualification of photomultipliers for use in neutrino telescopes~\cite{oldpmts,Aiello2025}. However, direct measurements of cesium vapor inside a sealed photomultiplier tube have never been attempted due to technical challenges. In this work, we present a noninvasive optical technique to detect cesium evaporation from an SbKCs photocathode during controlled heating of a PMT. By tuning a distributed feedback (DFB) laser across the cesium D\(_2\) resonance line at \SI{852.113}{\nano\meter}~\cite{Steck2019,Metcalf1999} and using differential absorption spectroscopy, we identify the temperature threshold and absorption strength associated with Cs release. The method is used to study the correlation between heating cycles and cesium vapor density, providing insight into the thermal dynamics of photocathode degradation.

\section*{Experimental Setup}
\begin{figure}[h!]
\centering
\begin{subfigure}[t]{0.59\textwidth}
\centering
\includegraphics[width=\textwidth]{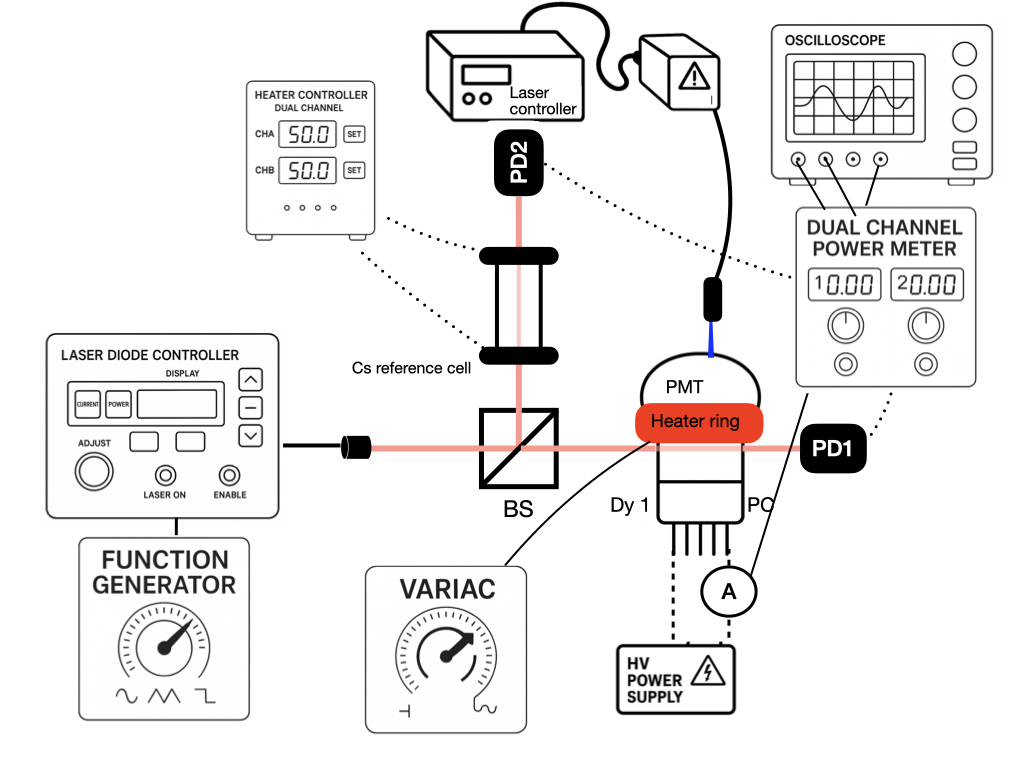}
\caption{Schematic diagram of the experimental setup. The DFB laser is split into two paths: a reference through a Cs vapor cell and a probe through the heated PMT. Additional paths include the pulsed laser excitation system and current measurement circuit.}
\label{fig:setup}
\end{subfigure}
\hfill
\begin{subfigure}[t]{0.39\textwidth}
\centering
\includegraphics[width=\textwidth]{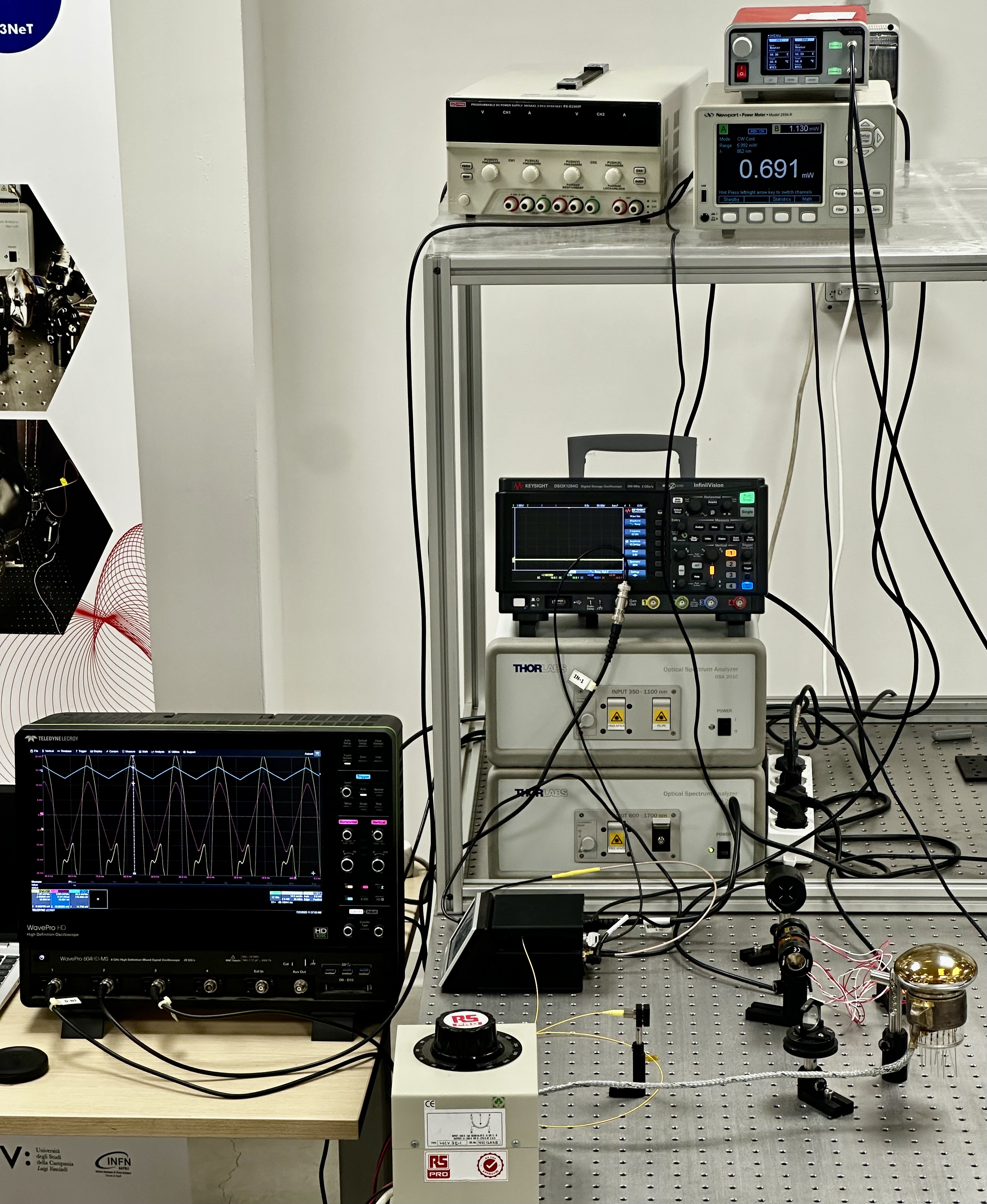}
\caption{Photograph of the experimental setup during a measurement.}
\label{fig:setup_photo}
\end{subfigure}
\caption{Overview of the experimental setup used for cesium desorption spectroscopy and photocurrent measurements.}
\label{fig:combined_setup}
\end{figure}
The experimental setup, schematically shown in Figure~\ref{fig:setup}, combines resonant laser absorption spectroscopy with photocurrent measurements under controlled thermal stress. A single-mode DFB laser diode (Thorlabs FLD LP852-SF60) is current-modulated through a Thorlabs CLD1010LP controller driven by a Keysight DSOX1204G oscilloscope in function generator mode. The modulation consists of a \SI{70}{\hertz} triangular waveform, enabling periodic tuning across approximately half of the cesium D\(_2\) hyperfine structure centered at \SI{852.1}{\nano\meter}. The output beam is divided by a 50:50 beam splitter: one branch traverses a temperature-stabilized cesium vapor reference cell (Thorlabs GC25075-CS) heated by two resistive heater rings (GCH 25R) under feedback control from a Thorlabs TC300B controller, while the other branch crosses the lower region of the PMT bulb, exploiting a free optical path between the internal dynode feedthroughs. Both transmitted beams are detected by Newport 918D-SL-OD3R photodetectors connected to a Newport 2936R dual-channel power meter, whose analog outputs are digitized by a LeCroy WaveRunner 604HD-MS oscilloscope (20~GS/s). The power meter was calibrated at \SI{852.1}{\nano\meter} using manufacturer-supplied correction factors, allowing absolute optical power determination with a residual uncertainty below 2\%. This ensured that transmission values could be converted into absorption coefficients and, through the Beer–Lambert relation, into in situ cesium vapor densities. The PMT under test is externally heated by a resistive heating band powered via a variac, which provides gradual and stable temperature ramps while minimizing thermal stress that could lead to bulb fracture. Real-time temperature distribution across the photocathode region is monitored with a calibrated FLIR T560 thermal camera, providing spatially resolved thermal maps synchronized with the optical data.  
In parallel, photocurrent response is characterized using a pulsed picosecond diode laser system (PILAS DX EGI-D2-40 with PiL1-040-40 head), emitting at either \SI{402}{\nano\meter} or \SI{509}{\nano\meter} depending on the measurement window. The source delivers \SI{100}{\pico\second} pulses at \SI{41}{\mega\hertz} and operates in pseudo continuum wave. The PMT is biased at \SI{-100}{\volt} by a CAEN DT5790M high-voltage power supply with the first dynode grounded, and the current between the photocathode and dynode is measured across the internal \SI{1}{\mega\ohm} input impedance of the same oscilloscope. This integrated configuration enables simultaneous acquisition of cesium absorption spectra and photocurrent traces during both heating and cooling cycles. The dual-diagnostic approach provides deterministic detection of cesium release through resonant absorption, while concurrently tracking quantum efficiency evolution via photocurrent response, thereby linking photocathode degradation mechanisms to in situ cesium dynamics.

\section*{Method}
This experiment was designed to induce controlled thermal stress on the bialkali photocathode and, for the first time, directly correlate cesium desorption with in situ optical absorption signatures and quantum efficiency changes. Absorption spectroscopy of the cesium D2 line at 852.1 nm has been widely employed in atomic physics and laser diagnostics~\cite{Steck2019,Metcalf1999}. The hyperfine-resolved structure and Doppler-broadened absorption profiles are well-documented~\cite{Corwin1998,Banerjee2001}, and various temperature-dependent studies have been conducted on cesium vapor~\cite{Zhang2010}. This work leverages these foundations to investigate cesium desorption from bialkali photocathodes via laser absorption through the photomultiplier envelope. A resistive heating band is wrapped around the PMT bulb and powered via a variac to incrementally raise the temperature of the photocathode. A distributed feedback (DFB) laser tuned around 852.1 nm is modulated with a current ramp to scan over the cesium D2 resonance line. The laser beam is divided into two branches: one is sent through a cesium vapor reference cell, while the second passes through the lower part of the PMT envelope. Both beams are detected by power meters and their analog outputs are digitized by an oscilloscope for real-time analysis. The comparison between absorption features in the reference cell and the PMT path provides a direct indication of cesium vapor release from the photocathode region.
Simultaneously, a pulsed picosecond laser diode (PiLas, 402 nm, 41 MHz repetition rate) is used to illuminate the photocathode in CW mode. The resulting photoelectron current is picked up between the photocathode (at -100,V) and the first dynode (at ground), across the \SI{1}{\mega\ohm} imput impedance of the oscilloscope. This configuration allows measuring both dark current and laser-stimulated current under the same thermal conditions by switching on and off the laser emission. The combination of absorption and current data permits assessing photocathode QE and correlating its degradation with cesium desorption.

\begin{figure}[h!]
    \centering
    \includegraphics[width=0.9\textwidth]{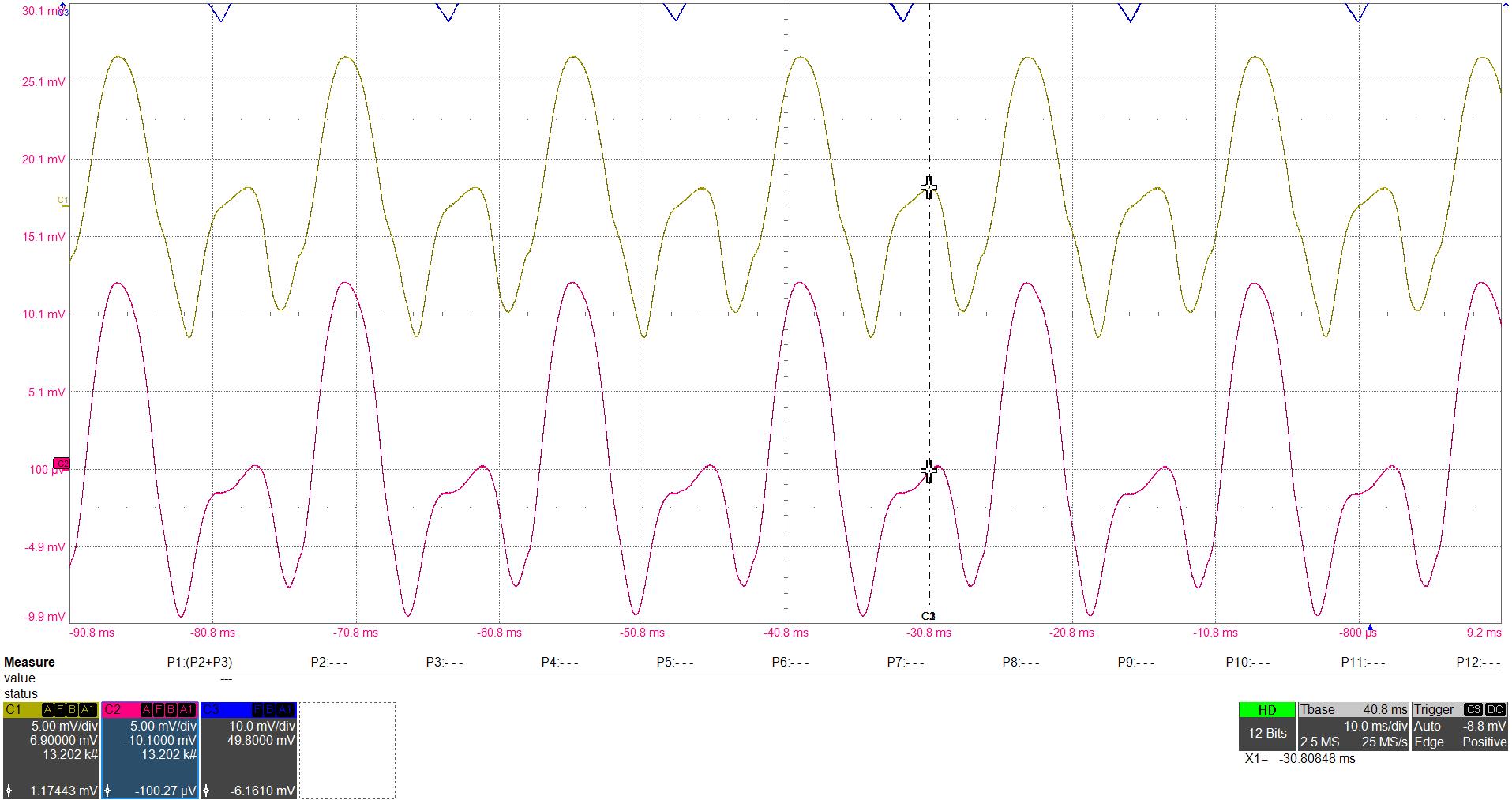}
    \caption{Representative absorption traces of the reference (Yellow) and probe (Pink) channels. The PMT signal reproduces the D2 line features only when the photocathode is heated beyond 60°C.}
    \label{fig:absorption}
\end{figure}

\subsection*{Estimation of Cesium Vapor Density}
The optical density of the observed absorption can be used to estimate the line-integrated cesium vapor concentration along the probe beam path. Assuming a Beer–Lambert absorption profile:
\[
\frac{I}{I_0} = \exp(-n \sigma L)
\]
where \( I/I_0 \) is the transmission ratio at line center, \( n \) is the cesium number density in \(\mathrm{cm}^{-3}\), \( \sigma \) is the absorption cross-section, and \( L \) is the optical path length through the vapor. For this experiment, we assume \( \sigma = 2.9 \times 10^{-13} \, \mathrm{cm}^2 \) and \( L = \SI{5}{\centi\meter} \). The analog output from the Newport 918D-UV power meter, digitized by the LeCroy oscilloscope, provides a measure of transmitted optical power. By comparing transmitted power at various temperatures to the baseline reference power (taken at \SI{24}{\celsius}), we estimated the relative absorption and derived the corresponding cesium number densities. 
The estimated uncertainty on the calculated cesium vapor densities is approximately \textbf{20\%}, primarily due to the combined contributions of the power meter calibration error (\(\sim 1{-}2\%\)), uncertainties in the absorption cross-section (\(\sim 5\%\)), and the estimated optical path length through the PMT envelope (\(\sim 4\%\)). The dominant factor is the sensitivity of the logarithmic dependence on transmission, particularly at high transmittance values.

\begin{figure}
    \centering
    \includegraphics[width=0.7\linewidth]{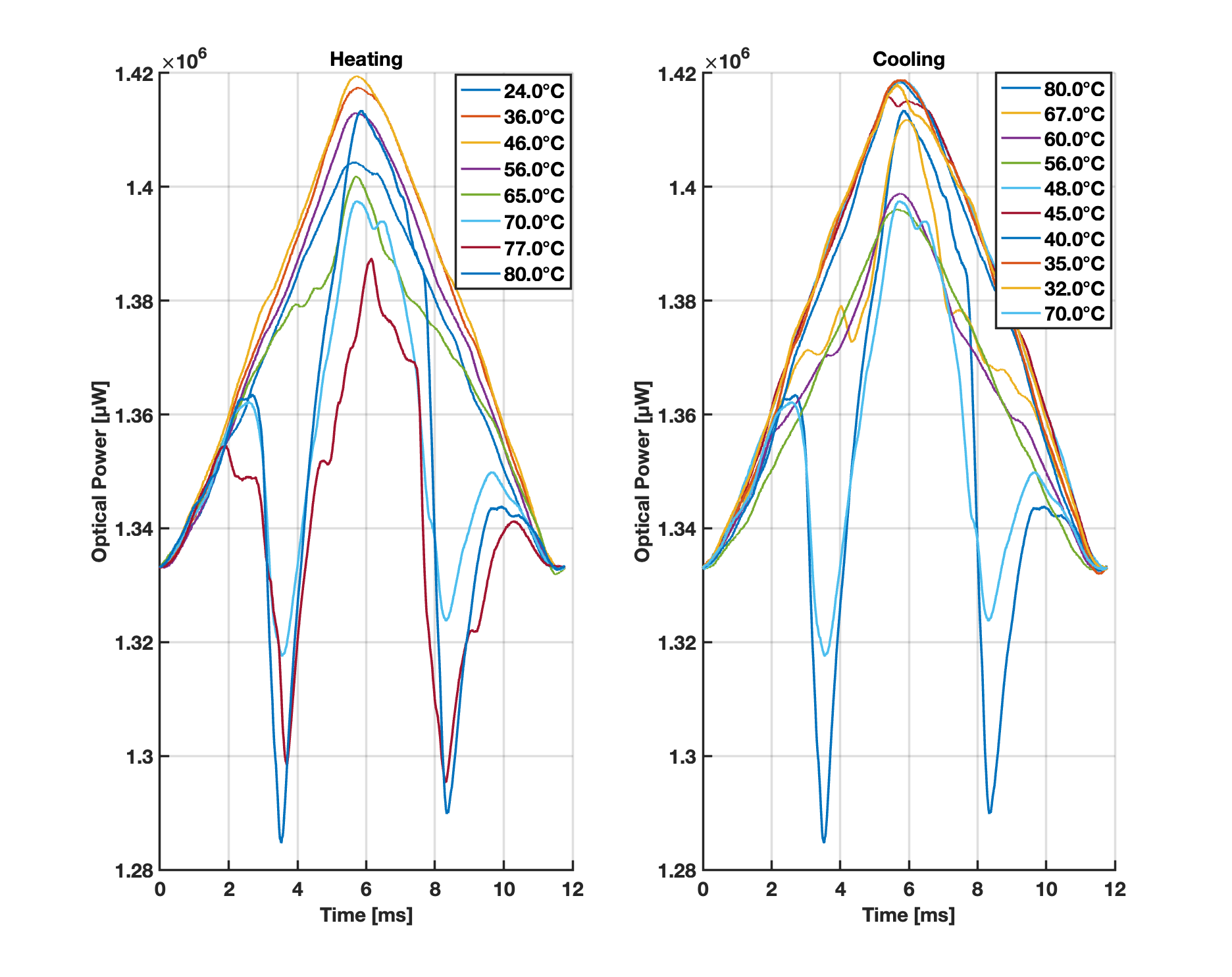}
    \caption{Stacking of 20 periods of the modulated signal}
    \label{fig:abs_evolution}
\end{figure}

The table \ref{tab:cs_density} summarizes the results for a set of selected temperatures, both in the heating and cooling phases:
\begin{table}[h!]
\centering
\caption{Cesium vapor density estimated via Beer–Lambert law at different photocathode temperatures.}
\begin{tabular}{cccc}
\hline
\textbf{T(°C)} & \textbf{Trans. Power \( I \) [\(\mu\)W]} & \textbf{Trans. Ratio \( I/I_0 \)} & \textbf{Density \( n \) [atoms/cm\(^3\)]} \\ 
\hline
80 (heating) & 469.00 & 0.9271 & \(5.36 \times 10^9\) \\
77 (heating) & 474.70 & 0.9380 & \(4.32 \times 10^9\) \\
70 (heating) & 481.00 & 0.9505 & \(3.38 \times 10^9\) \\
65 (heating) & 503.50 & 0.9949 & \(1.05 \times 10^8\) \\
80 (cooling) & 469.21 & 0.9275 & \(5.31 \times 10^9\) \\
70 (cooling) & 481.00 & 0.9505 & \(3.38 \times 10^9\) \\
67 (cooling) & 500.40 & 0.9890 & \(7.08 \times 10^8\) \\
\hline
\end{tabular}
\label{tab:cs_density}
\end{table}
The evolution of the absorption dip with temperature is further visualized in Figure~\ref{fig:abs_evolution}, where each line represents the stacking and averaging to obtain a single-period modulation ramp at a given temperature. 
As expected, cesium absorption increases with temperature due to enhanced thermal desorption, and decreases during cooling, confirming the reversible nature of Cs evaporation under thermal stress. This approach provides a quantitative and non-invasive method to probe the alkali vapor density in sealed photodetectors and demonstrates clear correlation with photocathode temperature, offering a diagnostic tool to assess the impact of thermal cycling on photocathode performance.

\section*{Results and discussion}
\begin{figure}[h!]
\centering
\begin{subfigure}[t]{0.49\textwidth}
    \centering
    \includegraphics[width=\linewidth]{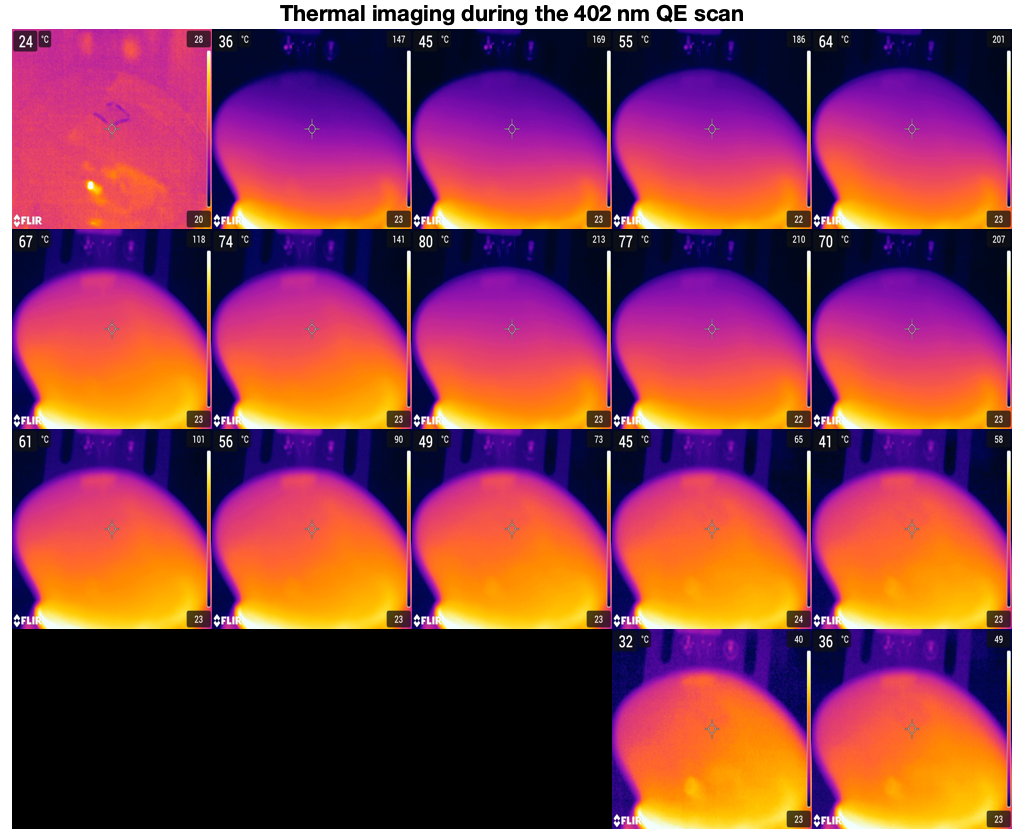}
    \caption{Thermal image of the PMT photocathode during the 402 nm scan.}
    \label{fig:thermal_402}
\end{subfigure}
\hfill
\begin{subfigure}[t]{0.49\textwidth}
    \centering
    \includegraphics[width=\linewidth]{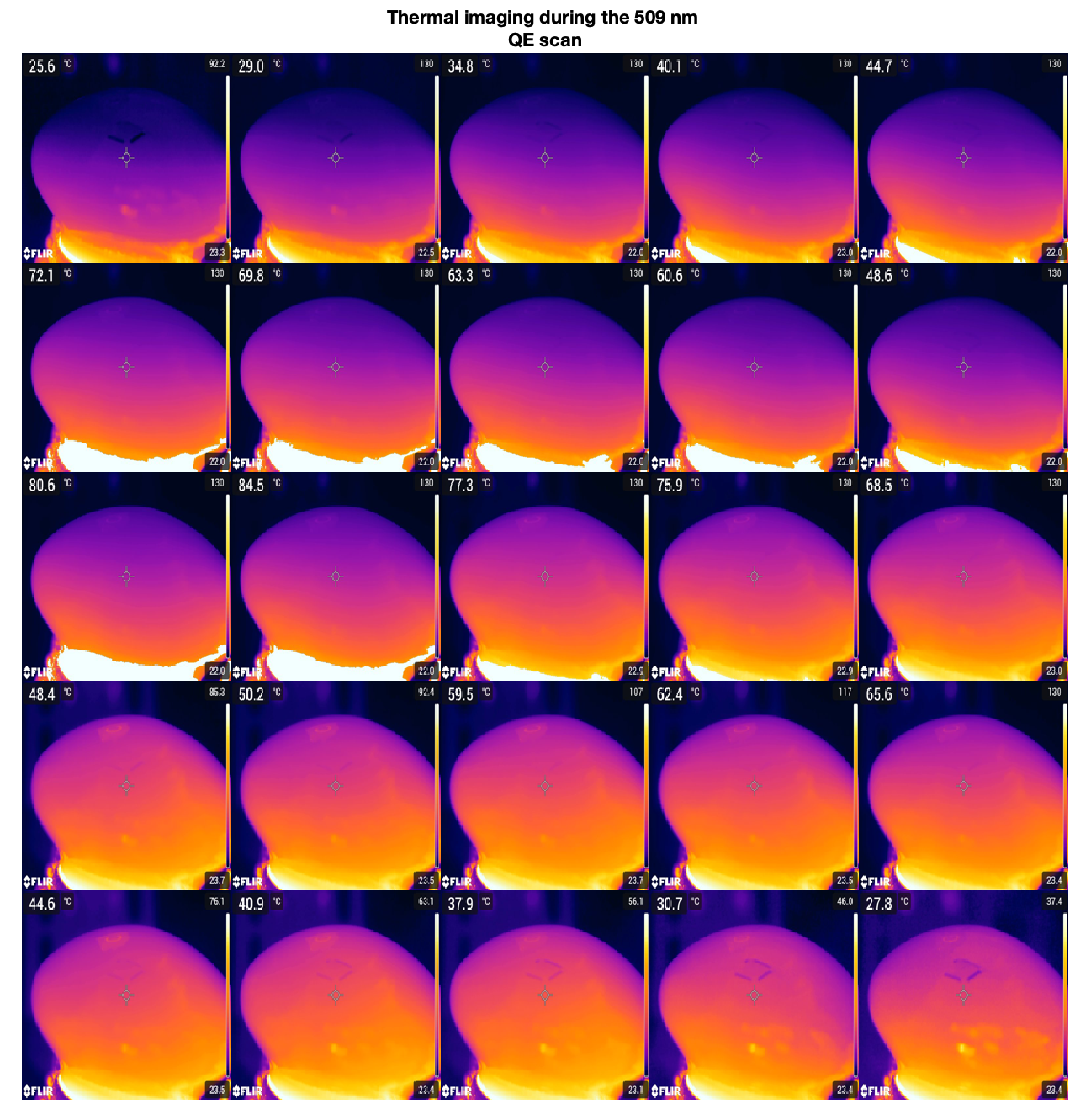}
    \caption{Thermal image of the PMT photocathode during the 509 nm scan.}
    \label{fig:thermal_509}
\end{subfigure}
\caption{Comparison of thermal conditions during the different wavelength scans. Both images were taken with a calibrated FLIR thermal camera targeting the photocathode area.}
\label{fig:thermal_comparison}
\end{figure}

\begin{figure}[htbp]
\centering
\rotatebox{90}{%
    \includegraphics[height=0.9\textwidth]{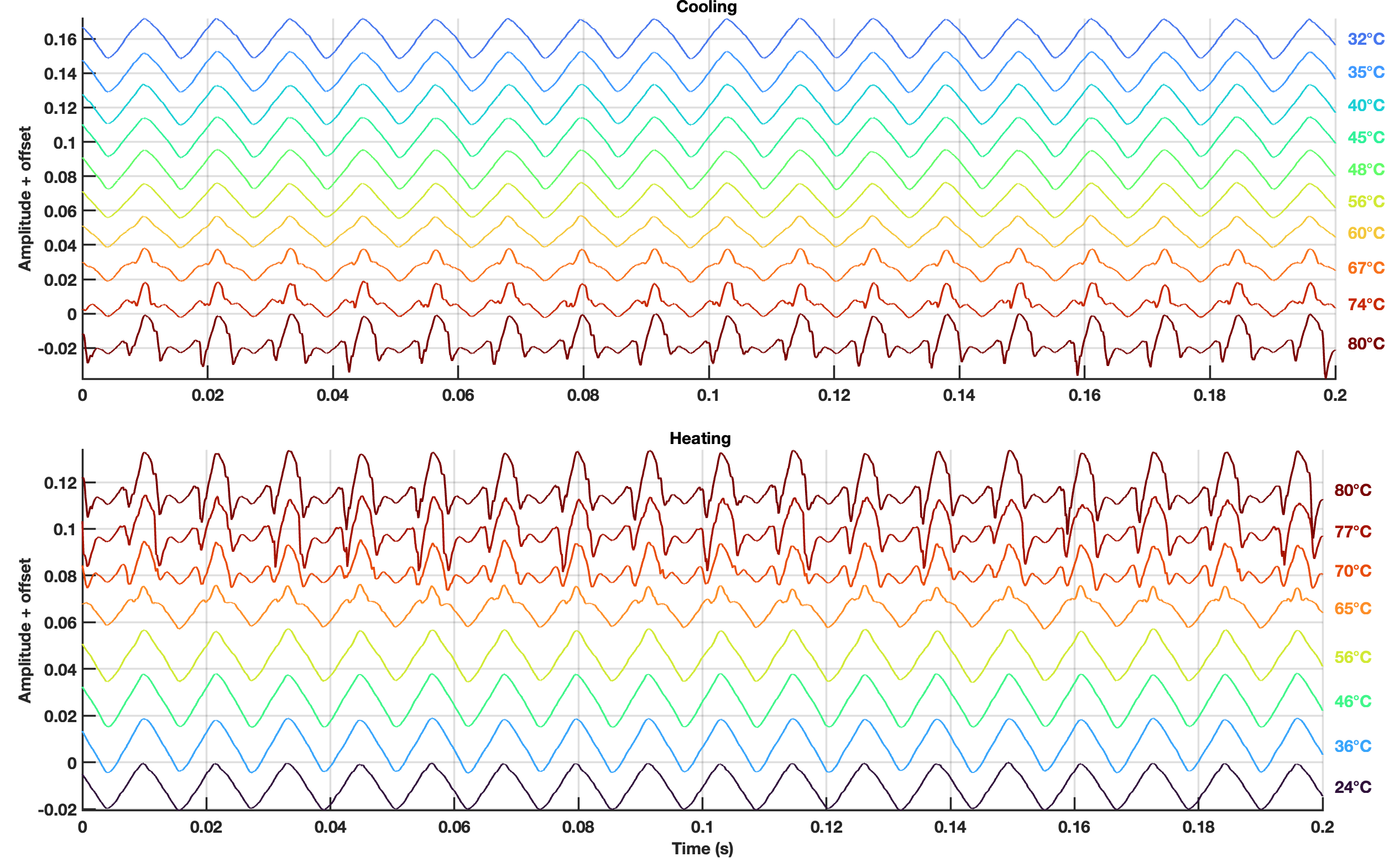}
}
\caption{Cesium absorption dips evolution during the QE scan at 402 nm}
\label{fig:abs_evolution402}
\end{figure}

\begin{figure}[htbp]
\centering
\rotatebox{90}{%
    \includegraphics[height=\textwidth]{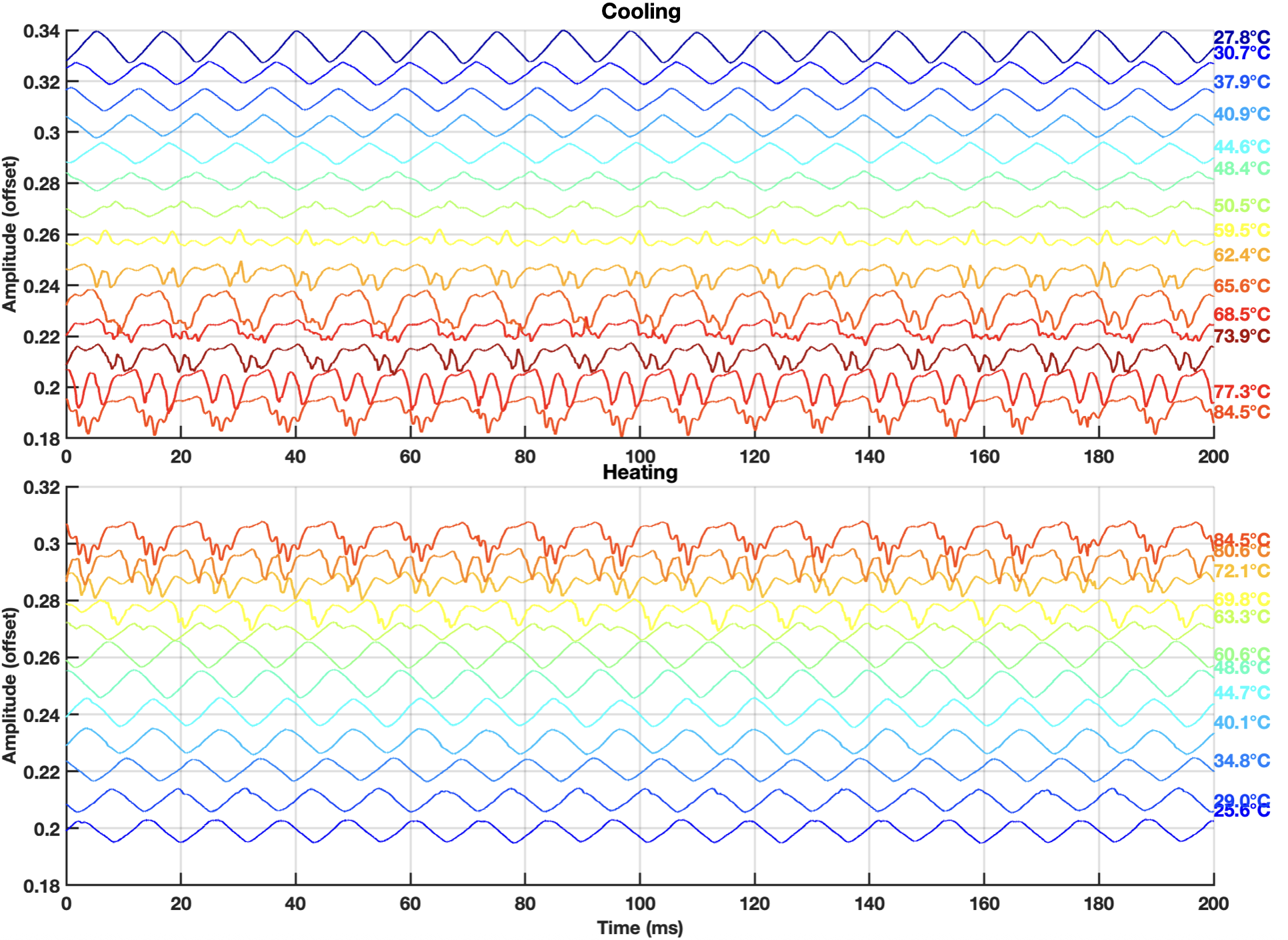}%
}
\caption{Cesium absorption dips evolution during the QE scan at 509 nm}
\label{fig:abs_evolution502}
\end{figure}

\begin{figure}[htbp]
\centering
\includegraphics[width=0.7\linewidth]{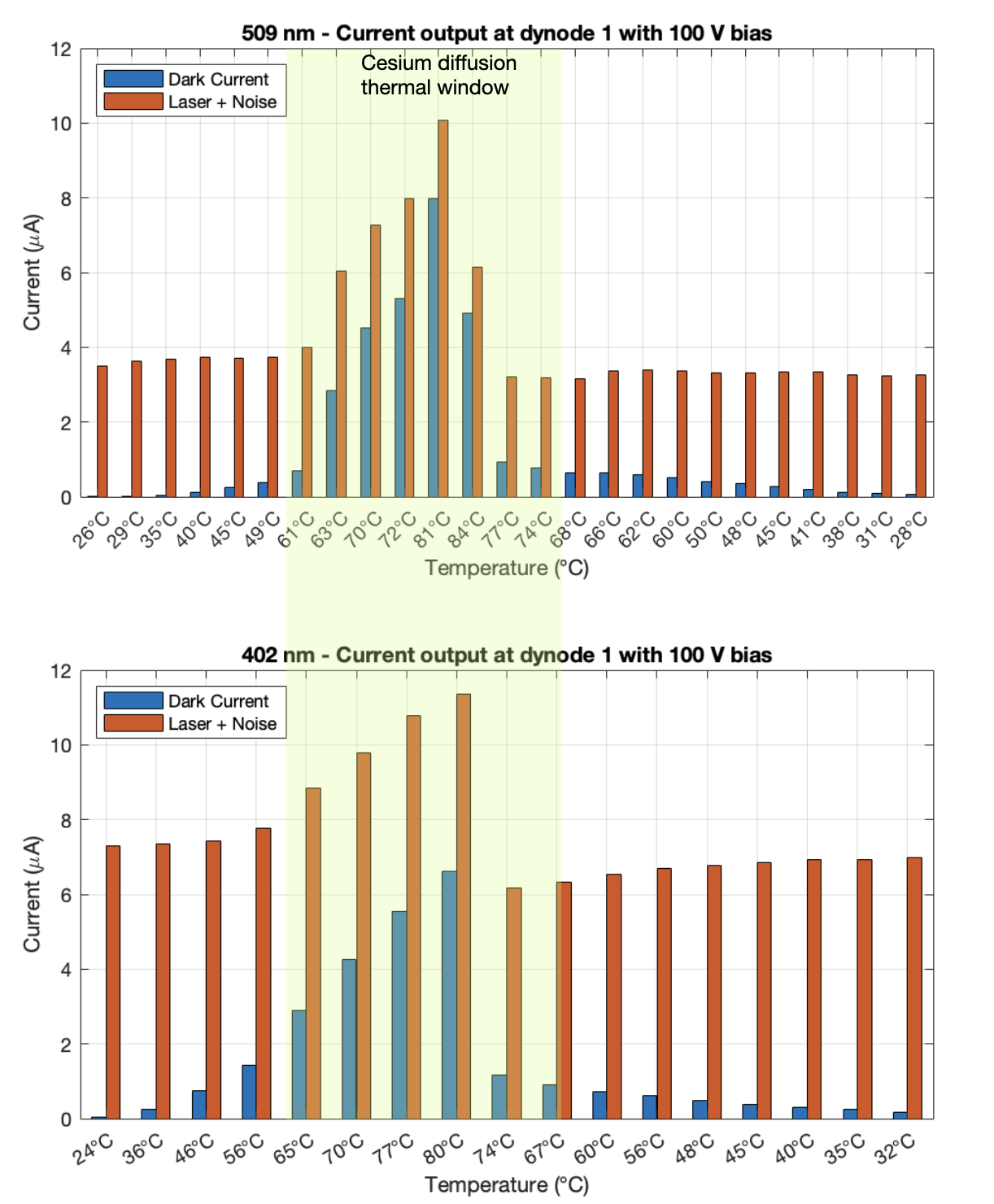}
\caption{Bar plot showing measured dark current and laser-stimulated current at each temperature during pulsed laser irradiation (402 nm and 509 nm, 41 MHz). The yellow shaded block highlights the observation window for cesium diffusion via absorption spectroscopy.}
\label{fig:bars_current}
\end{figure}

\begin{figure}[htbp]
\centering
\includegraphics[width=0.99\linewidth]{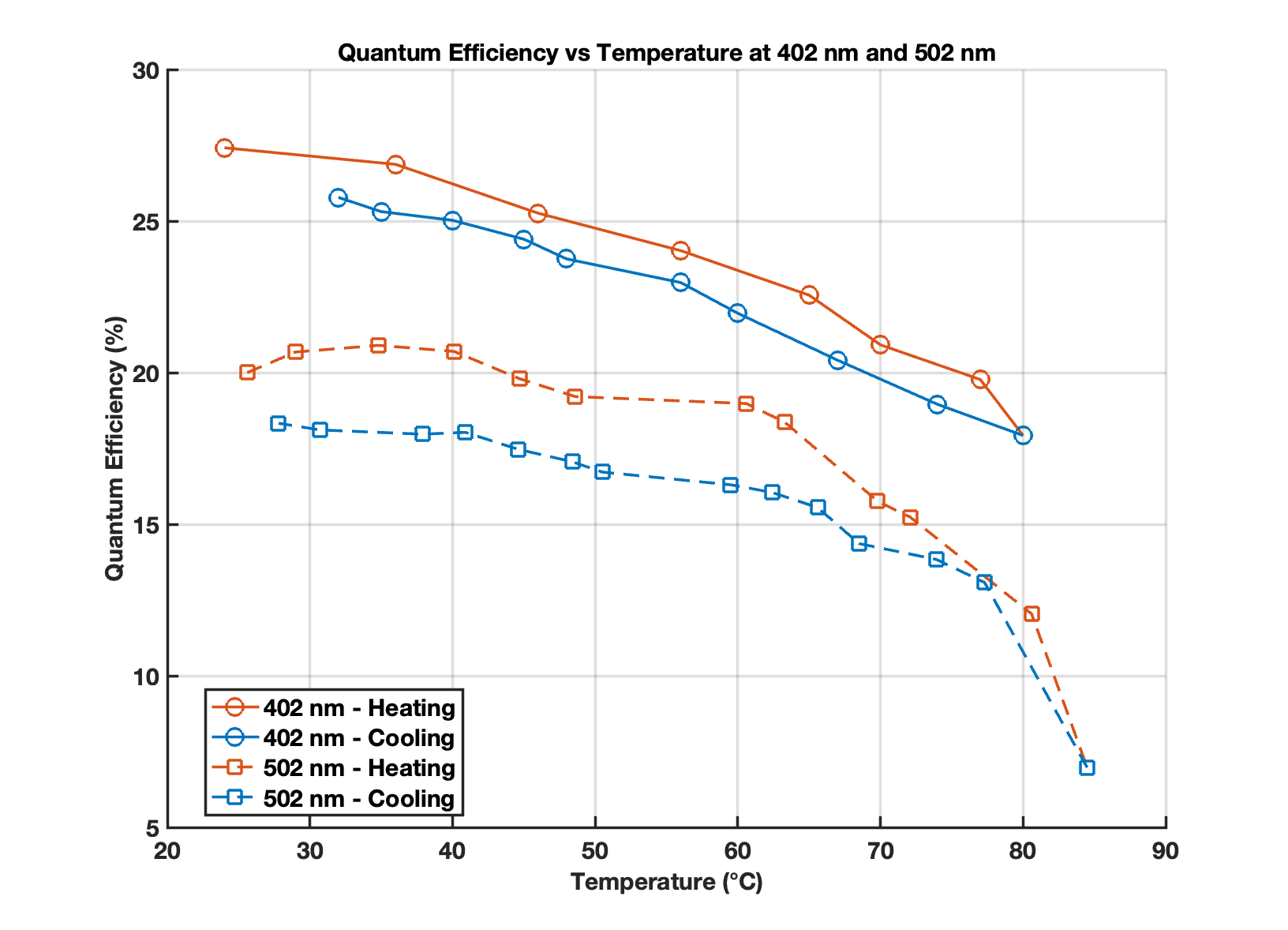}
\caption{Quantum efficiency (QE) evolution with photocathode temperature for wavelengths of 402 nm and 509 nm  Cs desorption causes a visible drop in QE which is partially reversible.}
\label{fig:qe_temp}
\end{figure}

For the \SI{402}{\nano\meter} measurements, data were acquired at 17 distinct photocathode temperatures between \SI{24}{\celsius} and \SI{80}{\celsius}, with \SI{200}{\milli\second} traces recorded at each step (Figure~\ref{fig:abs_evolution402}). At \SI{509}{\nano\meter}, the sampling was extended to 25 temperature points in order to provide a finer resolution of the thermal behaviour. For each temperature, measurements were carried out both with and without picosecond laser illumination (PiLas diode laser, \SI{402}{\nano\meter}, \SI{40}{\mega\hertz} repetition rate), allowing sequential acquisition of the dark current and the photocurrent. This dual acquisition was essential for extracting the quantum efficiency (QE), calculated as the ratio of emitted photoelectrons to incident photons. The oscilloscope traces from the reference cell and the PMT branch allow direct comparison of the cesium D\(_2\) absorption dip evolution with temperature. While the reference cell provided a stable spectroscopic fingerprint of the atomic resonance, the PMT branch displayed a temperature-dependent increase in absorption, revealing the progressive release of cesium vapor inside the sealed bulb. The acquisition was manually triggered to synchronize absorption spectra with photocurrent measurements. Figure~\ref{fig:bars_current} shows the current data at both wavelengths, where each temperature corresponds to two adjacent bars: one for dark current and one for photocurrent under laser illumination. A clear trend emerges: both dark current and photocurrent increase with rising temperature due to enhanced thermal activity and leakage currents within the PMT. However, the net photocurrent—defined as the difference between the illuminated and dark signals—progressively decreases as the temperature rises. This behaviour indicates a reduction in photoemission efficiency, consistent with quantum efficiency (QE) degradation. Notably, during the cooling phase, the net signal partially recovers, suggesting that the QE degradation is at least partially reversible. This trend supports the hypothesis that cesium desorption at elevated temperatures temporarily impairs the photocathode’s ability to emit electrons, without causing irreversible damage to the material or structure.
Figure~\ref{fig:qe_temp} reports the extracted QE values as a function of temperature for both heating and cooling cycles. The data reveal a reversible behavior: QE decreases during heating, recovers significantly during cooling, and closely follows the dynamics of cesium absorption detected spectroscopically. This represents the first direct, non-invasive, and real-time correlation of cesium vapor release with QE degradation inside a photomultiplier tube. By monitoring simultaneously the resonant absorption dip and the photocurrent response, it is possible to quantify both the cesium density within the bulb and the functional performance of the photocathode without damaging the device. The results demonstrate that cesium desorption under thermal stress is a key mechanism driving QE loss, but also that part of the degradation is reversible upon cooling, consistent with re-adsorption processes. The present findings provide experimental confirmation of the mechanism previously hypothesized in QE evolution studies under thermal stress in ~\cite{DeBenedittis2025}. Whereas earlier work suggested a link between alkali metal dynamics and photocathode performance, the current study establishes direct spectroscopic evidence of cesium release and correlates it with QE evolution. This dual-diagnostic method thus offers a new tool for assessing photocathode stability and predicting PMT performance under operating conditions.

\section*{Conclusions}
These results demonstrate that Cs desorption from bialkali photocathodes can be monitored in situ via resonant laser absorption, with detection thresholds on the order of $10^9$--$10^{10}$ atoms/cm$^3$. Additionally, concurrent photocurrent measurements under pulsed laser excitation reveal a clear reduction in QE at elevated temperatures, followed by partial recovery on cooling. This work establishes, for the first time, a spectroscopic correlation between Cs evaporation and photocathode QE loss. The method provides a precise and non-invasive diagnostic for understanding thermal degradation mechanisms in sealed photodetectors.

\end{document}